\documentclass[preprint]{aastex}

\newcommand{\be}{$^{7}$Be } 
\newcommand{\n}{$^{13}$N }

\newcommand{\oxoo}{$^{16}$O } 
\newcommand{\fl}{$^{17}$F } 
\newcommand{\f}{$^{18}$F } 
\newcommand{\na}{$^{22}$Na } 
\newcommand{\al}{$^{26}$Al } 
\newcommand{\msun}{M$_\odot$ }
\newcommand{\msyr}{M$_\odot$.yr$^{-1}$ } 

\begin{document}

\title{
Gamma-ray emission from novae related to positron annihilation: constraints 
on its observability posed by new experimental nuclear data
}

\author{Margarita Hernanz}
\affil{Institut d'Estudis Espacials de Catalunya/CSIC,
Edifici Nexus-201, C/Gran Capit\`a 2-4, E-08034 Barcelona, SPAIN}
\author{Jordi Jos\'e}
\affil{Departament de F\'{\i}sica i Enginyeria Nuclear (UPC), Avinguda
V\'{\i}ctor Balaguer, s/n, E-08800 Vilanova i la Geltr\'u (Barcelona),
SPAIN \\
and\\
Institut d'Estudis Espacials de Catalunya,
Edifici Nexus-201, C/Gran Capit\`a 2-4, E-08034 Barcelona, SPAIN}
\author{Alain Coc}
\affil{Centre de Spectrom\'etrie Nucl\'eaire et de Spectrom\'etrie de
Masse, IN2P3-CNRS, Universit\'e Paris Sud, B\^at.104, F-91405 Orsay Campus, 
FRANCE} 
\author{Jordi G\'omez-Gomar}
\affil{Institut d'Estudis Espacials de Catalunya,
Edifici Nexus-201, C/ Gran Capit\`a 2-4, E-08034 Barcelona, SPAIN}
\and
\author{Jordi Isern}
\affil{Institut d'Estudis Espacials de Catalunya/CSIC,
Edifici Nexus-201, C/Gran Capit\`a 2-4, E-08034 Barcelona, SPAIN}
	      
\received{}
\accepted{}

\slugcomment{\underline{Submitted to}: \apjl~~~~\underline{Version}:
\today}

\begin{abstract}
 
Classical novae emit $\gamma$-ray radiation at 511 keV and below, with a 
cut-off at around (20-30) keV,
related to positron annihilation and its Comptonization in the
expanding envelope. This emission has been elusive up to now, because it
occurs at epochs well before the maximum in optical luminosity, but it could
be detected by some sensitive intrument on board a satellite,
provided that the nova is close enough and that it is observed at the right 
moment. The detection of this emission, which is a 
challenge for the now available and for the future $\gamma$-ray instruments, 
would shed light into the physical processes occurring in the early phases of
the explosion, which are invisible in other lower energy ranges.
A good prediction of
the emitted fluxes and of the corresponding detectability distances with
different instruments relies critically on a good
knowledge of reaction rates relevant to \f destruction, which have been 
subject to a
strong revision after recent nuclear spectroscopy measurements.
With respect to previous results, smaller ejected masses of \f are predicted, 
leading to smaller emitted fluxes in the (20-511) keV range and shorter
detectability distances.

\end{abstract}


\section{Introduction} 

The role of novae as potential gamma-ray emitters has been mentioned long
ago (Clayton \& Hoyle 1974, Clayton 1981, Leising \& Clayton 1987), and 
analyzed in detail on the basis of theoretical models just
recently (G\'omez-Gomar et al. 1998). In many of the historical studies, more 
attention has
been paid to the long lasting emission, related to the decay of some 
medium-lived isotopes, such as \be ($\tau$=77days), \na ($\tau$=3.75yr), and 
to the possible contribution of novae to the Galactic content of \al 
($\tau=1.04 \times 10^6$ yr). 
Up to now, no positive detection neither of the 478 keV emission (Harris et al.
1991), related to
\be decay, nor of the 1275 keV one (Iyudin et al. 1995), related to \na decay, 
has been obtained. The 1809 keV line associated with 
the galactic \al was detected some
years ago by the HEAO3, High Energy Astrophysics Observatory 
(Mahoney et al. 1982), but observations made with 
the COMPTEL (Compton Telescope) instrument on board the CGRO satellite 
(Compton Gamma-Ray Observatory) seem to indicate that the \al emission
is more related to a young population of massive stars, although a
nova contribution cannot be ruled out yet (Diehl et al. 1995, Prantzos \& 
Diehl, 1996). The most powerful emission
in gamma-rays from classical novae is not the one related to the 478, 1275
or 1809 keV lines, but that at 511 keV and below (down to $\sim$(20-30) keV), 
which
originates from electron-positron annihilation (Hernanz et al. 1997a, b, 
G\'omez-Gomar et al. 1998), and consists of a line at 511 keV and a continuum
below it. The 511 keV line comes from the direct
annihilation of positrons and from the positronium (in singlet state) 
emission, whereas the continuum comes from both the positronium continuum
(triplet state positronium) and the Comptonization of photons emitted in the
511 keV line. The main contributors to positrons in nova envelopes are the 
short-lived \n ($\tau$=862s) and \f ($\tau$=158 min), besides the long-lived 
\na (this one only in oxygen-neon novae). 
The problem with this emission is its very short duration, making its 
detection very difficult. Furthermore, because of the short lifetimes of 
\n and \f, its maximum happens well before the maximum in
visual light, which still enhances the difficulty of detection. 
When \n decays, the nova envelope 
is still too opaque; therefore, \f plays 
the most important role, since it decays appreciably when the envelope begins 
to be transparent enough to allow for the gamma-ray photons to be transported
through the whole envelope, and then be emitted into space. 

Several attempts are being carried out to detect the annihilation gamma-ray 
emission from novae, with instruments like BATSE (Burst and Transient 
Source Experiment) on board the CGRO, or TGRS (Transient Gamma Ray 
Spectrometer), on board the WIND satellite. SPI, the future 
spectrometrer for INTEGRAL (International Gamma-Ray Laboratory), 
could in favorable conditions detect that emission. 
The detectability distances for 
these instruments, or for any future ones, rely on the detailed
properties of novae expanding envelopes and, in particular, on their 
\f content,
which strongly depends on the rates of the nuclear reactions involved in its
synthesis. 
A good knowledge of the gamma-ray output of novae is needed to elucidate the 
possibilities of detection and to put constraints on the theoretical models 
from the observations. 

\section{Gamma-ray emission of novae and \f synthesis}

The flux emitted by a classical nova in the gamma-ray range below 511 keV
depends on the amount of \f at the epoch where the envelope begins to
be transparent to gamma-rays. Therefore, profiles of \f along the
expanding envelope at different times are needed to compute the spectrum at 
different epochs after the outburst (or in an equivalent way, the light curves 
for different gamma-ray bands at and below 511 keV). Of course, 
the whole spectrum depends also on the abundances of the other isotopes  
and on the density and velocity profiles; these quantities determine the 
number of 
interactions of photons with the ejecta through Compton scattering,
photoelectric absorption and pair production. 
Theoretical models of the gamma-ray emission from individual classical novae, 
based on the nucleosynthetic yields obtained
from a hydrodynamical code that follows the explosive phase and the related 
nucleosynthesis (Jos\'e \& Hernanz 1998), predict an intense line emission 
at 511 keV plus a continuum below (G\'omez-Gomar et al. 1998). The emission 
appears between 5 and 6 hours after maximum temperature, i.e., some days
before the maximum in visual luminosity. The continuum displays always a
cut-off at low energies, which is related to photoelectric absorption, which
has larger cross sections than Compton scattering at low energy, acting as a
sink of the Comptonized photons. The cut-off is located at $\sim$20 keV, for 
the carbon-oxygen (CO) and $\sim$30 keV for the oxygen-neon (ONe) rich 
envelopes typical of nova models. 

The main nuclear path leading to \f synthesis in novae belongs to the hot CNO 
(carbon-nitrogen-oxygen) cycles. Since in both CO and ONe 
novae the initial abundance of \oxoo is large, \oxoo is the main 
source for \f formation, through two possible chains of reactions: 
\oxoo(p,$\gamma$)\fl(p,$\gamma$)$^{18}$Ne($\beta^+$)\f and 
\oxoo(p,$\gamma$)\fl($\beta^+$)$^{17}$O(p,$\gamma$)\f.
However, \f yields are severely constrained by its destruction mode, 
whatever the production channel is.
During the thermonuclear runaway, \f destruction by beta decays can be 
neglected
when compared to its destruction by proton captures. Because of the low alpha 
emission threshold, \f(p,$\alpha$)$^{15}$O is faster than 
\f(p,$\gamma$)$^{19}$Ne, and hence it is the main destruction channel of \f 
(see Figure 1). 
The rates for \f(p,$\gamma$) and \f(p,$\alpha$) generally adopted in nova
models are those from Wiescher \& Kettner (1982), which were estimated at a 
time where the relevant experimental data were scarce (i.e.,   
very limited spectroscopic data was available for $^{19}$Ne).  
Thanks to recent measurements, in 
particular to those of Graulich et al. (1997), Utku et al. (1998), and
references therein, it is possible now to use updated rates.
Two resonance strengths have been measured directly (Coszach et al. 1995, 
Graulich et al. 1997) in 
\f(p,$\alpha$) using a radioactive \f\ beam. Of these two, the one lying at 
higher energy is so broad that its tail alone gives a contribution larger 
than the Wiescher \& Kettner (1982) rate at nova temperatures.
Nuclear spectroscopic studies performed by Utku et al. (1998) have led to the
localisation of several new $^{19}$Ne levels to which should correspond new
resonances in \f+p. 
They have calculated updated \f(p,$\gamma$) and \f(p,$\alpha$)
rates including the two directly measured resonances plus the contribution of
the expected ones. To calculate this last contribution, they used
information about $^{19}$F analog levels supplemented with estimated
quantities.
A comparison of the new rates with those of Wiescher and Kettner is shown 
in Figure 1;  
in the range of temperatures relevant to nova explosions, the new rates are
quite larger than the older ones. 
Although the uncertainties associated with the rates of \f destruction, 
\f(p,$\gamma$) and \f(p,$\alpha$), are far from being settled in the nova
temperature range, it seems clear
that the rates that had been used up to now (Wiescher \& Kettner, 1982) 
were underestimated. 

In Table 1 we show the most relevant properties of some nova models computed 
by means of the hydrodynamic code described in Jos\'e \& Hernanz (1998), 
with the nuclear reaction network described in Jos\'e et al. (1999)
(recommended rates) and 
the new Utku's rates for \f(p,$\gamma$) and \f(p,$\alpha$) reactions. 
We have chosen models representative of the two nova types, CO and ONe, and 
with
masses 1.15 \msun in the CO case and 1.15 and 1.25 \msun in the ONe case, in
order to analyze the influence of the composition and mass of the underlying
white dwarf on \f synthesis. 
The accretion rate in all the cases is $2 \times 10^{-10}$ \msyr and the
degree of mixing with the core is 50\% (see Jos\'e \& Hernanz 1998 for a
discussion of the initial mixing effect). 
Similar ejected masses of \n and \f are obtained in all the models, which is
not the case for other radioactive nuclei relevant to $\gamma$-ray emission
with longer timescales,
such as \be (mainly produced in CO novae) and \na and \al (mainly synthesized
in ONe novae). 
When compared with results with the old rates for \f+p reactions, a general
reduction, by a factor between 10 and 20, of \f yields is present, whereas 
the maximum temperature and the mean kinetic energy are almost
unchanged, since the \f+p reactions do not affect the energetics of the nova
explosion. Concerning the general nucleosynthesis, 
the changes are not very relevant, except for $^{19}$F, which is reduced 
typically by a factor of 10. 

The gamma-ray spectrum 12 hours after peak temperature is shown in Figure 2, 
for a CO nova of 1.15 \msun and an ONe nova of 1.25 \msun at a distance of 
1 kpc. 
The 511 keV line from our models is slightly blueshifted (with a maximum 
energy of 517 keV) during the first phases, and has a moderate width (from 
$\sim$5 to $\sim$8 keV, FWHM, depending on the model and on the epoch).
Light curves in the 20-511 keV band and in the 511 keV line alone 
have been computed. A summary of their fluxes at different 
epochs after peak temperature, for a distance of 1 kpc, is shown in Table 2. 
There is an early maximum at around 1 hour, which is related to \n decay. 
This maximum has very short duration and is very dependent on the 
distribution of \n in the outer layers of the envelope (which are the only 
ones seen at these early epochs). A larger abundance of \n in the outer layers 
of the CO nova with respect to the ONe one leads to the larger flux in the 
CO case (see Table 2).
The flux reaches its second and broader maximum, related to \f decay, 
at around 6 hours after peak temperature.
It is important to notice that the flux emitted in the continuum, with
a maximum around 50 keV, is larger than that in the line at all times. 
Also, the clear cut-off at $\sim$20 keV, together with the short duration of 
the whole emission and its early
appearence before maximum visual luminosity, rules out the interpretation of
the early hard X-ray emission observations of Nova Herculis 1991 (Lloyd et al. 
1992), and of the recent Nova Velorum 1999 (Orio et al. 1999, 
Mukai \& Ishida 1999),
as a Comptonization of the gamma-ray emission 
(Livio et al. 1992, Starrfield et al. 1992). For the purpose of comparison,
the gamma-ray spectrum of the CO nova model, but computed with the old \f+p
rates, is also shown in Figure 2. 
A reduction by a factor of $\sim$10 at all the energies in the relevant 
range is obtained in the new models, since 
the fluxes are smaller than the previous ones by the same factors than 
\f ejected masses (between 10 and 20). This difference appears during the first 
24 hours, when \f is the main
source of positrons; but later on, when only \na 
remains as a positron source, the new and the old fluxes are quite similar,
being much smaller in CO novae because they are 
almost devoid of \na.

\section{Discussion and conclusions}

The potential detection of the electron-positron annihilation emission from
classical novae is faced with various challenging problems: besides of the 
relatively small fluxes, the short duration and the early appearence of the
emission make its detection quite hard. Large field of view intruments, like
BATSE, are the more appropriate to detect this type of emission. Its
capability to observe all the sky, together with its high sensitivity in the
low-energy range make BATSE an ideal instrument to detect this emission,
although the new fluxes presented here restrict the detectability distances to
less than 3 kpc. There is an ongoing project of analysis of BATSE 
background data for nearby novae that have exploded since CGRO launch.

Another instrument which could be able to detect the annihilation gamma-ray 
emission from novae, because of both its wide field of view and
its high spectral resolution, is the Transient Gamma Ray Spectrometer (TGRS)
on board the WIND satellite (Teegarden et al. 1996). 
Very recently, Harris et al. (1999) have tried to detect 
the annihilation line emission at 511 keV from the five
known Galactic novae in the instrument field of
view, during the time interval 1995-1997. No definite detections were made,
but the method applied was shown to be
sensitive enough to detect novae out to about 2.5 kpc. Since the conclusions 
of that paper were based on models in G\'omez-Gomar et al. (1998), 
they should be revised and the corresponding distances should be
reduced by a factor $\sim \sqrt {10}$. 

Finally, we should mention the future spectrometer for INTEGRAL, SPI, which
could be able to detect the 511 keV line up to 3 kpc with the fluxes we
obtain now. A similar detectability distance is predicted for the continuum. 
It is worth noticing that both TGRS and SPI have very good spectral
resolution, because of their germanium detectors; this of course is favorable 
for
the detection of the 511 keV line (although its non negligible width worsens
the sensitivity appreciably) but not so interesting for the continuum, where
novae emit a larger flux. 

The detection by some present or future instrument of the positron 
annihilation emission from novae would help to understand the mechanism of 
the explosion and the dynamics of the early expansion stage.  

\acknowledgments
This research has been partially supported by the CICYT-P.N.I.E.
(ESP98-1348), by the DGICYT (PB97-0983-C03-02; PB97-0983-C03-03), by the
CIRIT (GRQ94-8001) and by the PICS 319.

\clearpage

\figcaption[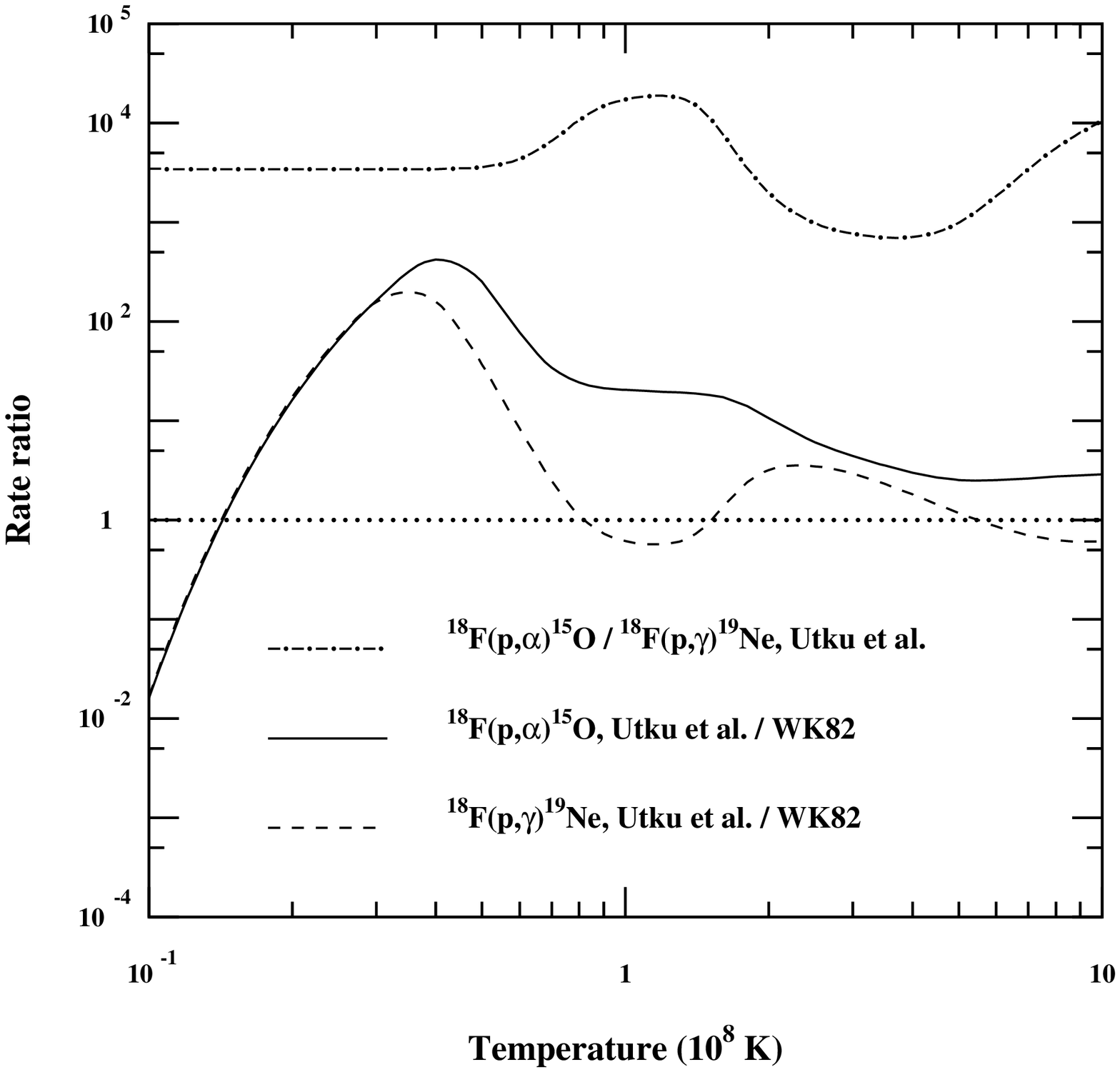]{Comparison between the Utku et al. (1998) rates and those
from Wiescher \& Kettner (1982), as well as comparison between \f(p,$\alpha$)
and \f(p,$\gamma$) rates, for a wide temperature range.}

\figcaption[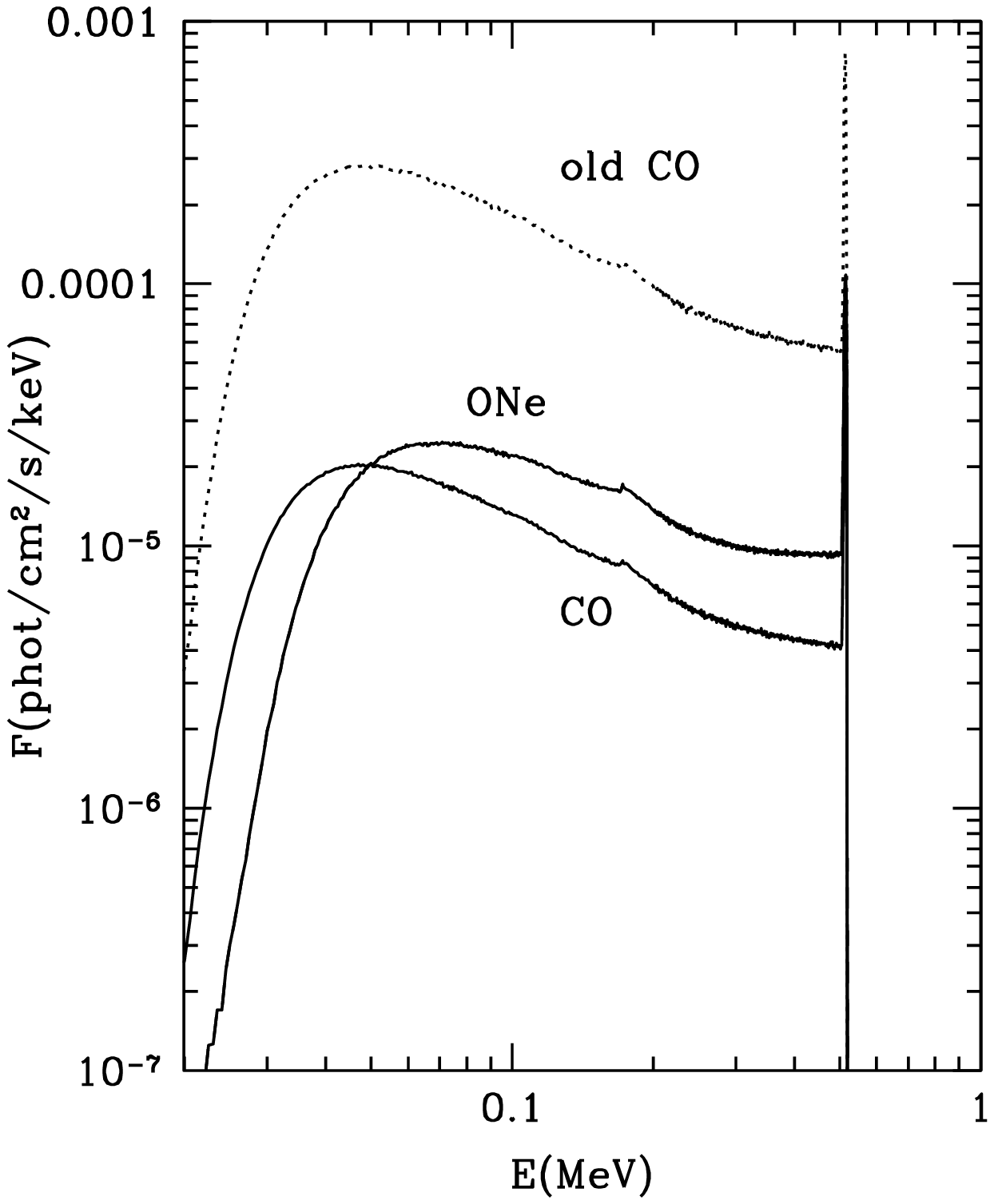]{Gamma-ray spectrum at t=12 h after peak temperature,
for a CO nova of 1.15 \msun (solid line) and an ONe nova (dashed line).
The old spectrum from G\'omez-Gomar et al. (1998) for the CO nova model is 
also shown for comparison (dotted line).}

\clearpage

\begin{table}
\begin{center}
\caption{Main properties of the ejecta 1 hour after peak temperature. 
T$_{\rm max}$ is the peak temperature, $<E_{\rm kin}>$ the mean kinetic energy, 
in units of $10^{45}$ ergs. The ejected masses of \n and \f are in \msun.}
\begin{tabular}{cccccc} 
\tableline\tableline
Nova type & M$_{\rm wd}$ (\msun) & T$_{\rm max}(K)$     & $<E_{\rm kin}>$ & 
            M$_{\rm ej}$(\n)     & M$_{\rm ej}$(\f)           \\
\hline
CO        & 1.15                 & $2.05 \times 10^{8}$ &  1.1  &
            $2.3 \times 10^{-8}$ & $2.6 \times 10^{-9}$ \\
ONe       & 1.15                 & $2.31 \times 10^{8}$ &  1.5  & 
            $2.9 \times 10^{-8}$ & $5.9 \times 10^{-9}$ \\
ONe       & 1.25                 & $2.51 \times 10^{8}$ &  1.5  &
            $3.8 \times 10^{-8}$ & $4.5 \times 10^{-9}$ \\
\tableline
\end{tabular}
\end{center}
\end{table} 

\begin{table}
\begin{center}
\caption{Fluxes, in photons.s$^{-1}$.cm$^{-2}$, at different 
times after peak 
temperature, in two energy bands in the continuum and in the 511 keV line
(FWHM=8 keV), for a CO nova of 1.15 \msun and an ONe nova of 1.25 \msun, 
at a distance of 1 kpc. The fraction of escaping energy in gamma-ray photons
is displayed in the last column.}
\begin{tabular}{cccccc} 
\tableline\tableline
t(hours) & Model 
         & F$_{(20-250) {\rm keV}}$   & F$_{(250-511) {\rm keV}}$ 
         & F$_{511 {\rm keV~line}}$   & fraction escaping (in \%)\\
\hline
1        & CO             
         & $2.5 \times 10^{-1}$       & $9.6 \times 10^{-2}$ 
         & $2.2 \times 10^{-2}$       & 0.4\\
         & ONe            
         & $2.2 \times 10^{-2}$       & $9.3 \times 10^{-3}$ 
         & $1.8 \times 10^{-3}$       & 0.02\\
\hline
3        & CO             
         & $5.3 \times 10^{-3}$       & $1.8 \times 10^{-3}$ 
         & $3.6 \times 10^{-4}$       & 1.9\\
         & ONe            
         & $1.6 \times 10^{-2}$       & $7.2 \times 10^{-3}$ 
         & $1.6 \times 10^{-3}$       & 4.2\\
\hline
6        & CO    
         & $8.1 \times 10^{-3}$       & $2.9 \times 10^{-3}$ 
         & $6.0 \times 10^{-4}$       & 10\\
         & ONe   
         & $1.7 \times 10^{-2}$       & $8.8 \times 10^{-3}$ 
         & $1.9 \times 10^{-3}$       & 17\\
\hline
12       & CO    
         & $2.5 \times 10^{-3}$       & $1.2 \times 10^{-3}$ 
         & $3.0 \times 10^{-4}$       & 39\\
         & ONe   
         & $3.8 \times 10^{-3}$       & $2.6 \times 10^{-3}$ 
         & $6.7 \times 10^{-4}$       & 47\\
\hline
24       & CO   
         & $2.5 \times 10^{-5}$       & $3.0 \times 10^{-5}$ 
         & $8.4 \times 10^{-6}$       & 77\\
         & ONe 
         & $5.2 \times 10^{-5}$       & $7.4 \times 10^{-5}$ 
         & $2.0 \times 10^{-5}$       & 80\\
\hline
48       & CO  
         & $1.5 \times 10^{-7}$       & $2.3 \times 10^{-6}$ 
         & $2.0 \times 10^{-8}$       & 95\\
         & ONe  
         & $1.1 \times 10^{-5}$       & $2.7 \times 10^{-5}$ 
         & $8.2 \times 10^{-6}$       & 94\\
\tableline
\end{tabular}
\end{center}
\end{table} 

\end{document}